# Implementing multiple imputation for missing data in longitudinal studies when models are not feasible: A tutorial on the random hot deck approach

Chinchin Wang[a,b], Tyrel Stokes[c], Russell Steele[c], Niels Wedderkopp[d], Ian Shrier[a]

[a] Centre for Clinical Epidemiology, Lady Davis Institute, Jewish General Hospital, McGill University, 3755 Côte Ste-Catherine Road, Montreal, Quebec, Canada H3T 1E2

[b] Department of Epidemiology, Biostatistics and Occupational Health, McGill University, 1020 Pine Avenue West, Montreal, Quebec, Canada H3A 1A2

[c] Department of Mathematics and Statistics, McGill University, 805 Sherbrooke Street West, Montreal, Quebec, Canada H3A 0B9

[d] Orthopedic Department University Hospital of South West Denmark, Department of Regional Health Research, University of Southern Denmark

**Corresponding Author:**

Ian Shrier MD, PhD

Centre for Clinical Epidemiology, Lady Davis Institute, Jewish General Hospital, McGill University, 3755 Côte Ste-Catherine Road, Montreal, Quebec, Canada H3T 1E2

Email: ian.shrier@mcgill.ca

Phone Number: 1-514-229-0114

# Abstract

**Objective:** Researchers often use model-based multiple imputation to handle missing at random data to minimize bias while making the best use of all available data. However, there are sometimes constraints within the data that make model-based imputation difficult and may result in implausible values. In these contexts, we describe how to use random hot deck imputation to allow for plausible multiple imputation in longitudinal studies.

**Study Design and Setting:** We illustrate random hot deck multiple imputation using The Childhood Health, Activity, and Motor Performance School Study Denmark (CHAMPS-DK), a prospective cohort study that measured weekly sports participation for 1700 Danish schoolchildren. We matched records with missing data to several observed records, generated probabilities for matched records using observed data, and sampled from these records based on the probability of each occurring. Because imputed values are generated randomly, multiple complete datasets can be created and analyzed similar to model-based multiple imputation.

**Conclusion:** Multiple imputation using random hot deck imputation is an alternative method when model-based approaches are infeasible, specifically where there are constraints within and between covariates.

**Keywords:** Multiple imputation, missing data, missing at random, hot deck imputation, random hot deck imputation, longitudinal studies

# Introduction

Missing data are present in most epidemiologic studies.[1] How they are handled is critical, as results may be biased and precision overestimated if inappropriate methods are used.[2] This tutorial article illustrates the use of random hot deck imputation to multiply impute unbiased values with appropriate confidence interval coverage when model-based methods are infeasible due to constraints between variables. Our example uses sport participation data from the Childhood Health, Activity, and Motor Performance School Study Denmark (CHAMPS-DK)[3] containing multiple constraints between variables. For instance, activity frequency must greater than or equal to the number of sports played, and individual sport frequencies must equal the total activity frequency. This approach is applicable to any context with important constraints between variables (e.g. side effects constrained by drug types; symptoms constrained by disease).

# Overview of imputation

Appropriate methods for imputation depend on the nature of missingness. Missing data fall into three categories: missing completely at random (MCAR), missing at random (MAR), and missing not at random (MNAR). Data are MCAR when missingness is independent of both observed and unobserved variables. Data are MAR when the missingness is only associated with observed variables. Data are MNAR when missingness is associated with unobserved variables.[4]

A simple way to handle missing data is to only analyze entries with complete data, i.e. complete case analysis. Complete case analysis is generally only unbiased when data are MCAR.[5] MCAR is often an unreasonable assumption due to factors associated with both missingness and the outcome.[6] For instance, data collection might be more difficult in participants who are sicker and more likely to die.

Researchers typically work under the weaker assumption that data are MAR. Established methods to handle MAR data without bias generally fall under model-based imputation, where models are used to predict missing values based on observed data.[7,8] MNAR data require advanced methods with complex assumptions, and are beyond the scope of this article.[4]

In addition to bias, researchers should account for imputed values having increased uncertainty compared to observed values.[9] Single imputation methods impute one replacement value for each missing value. They do not account for uncertainty in the imputed value resulting in confidence intervals that are too narrow.[8] Common single imputation methods include last observation carried forward[10] and mean imputation.[11] Multiple imputation methods account for increased uncertainty by imputing with random variation to create several complete datasets. Each dataset is analyzed and results are averaged,[8] providing more appropriate confidence intervals than single imputation.[9]

Multiple imputation is commonly applied to model-based techniques. Briefly, a model is generated that predicts missing values by borrowing information on observed relationships between covariates in the data.[7,8] Regression parameters and/or imputed values are sampled from an underlying distribution[12] to generate different imputed values in multiple different data sets. This is generally straightforward for continuous data, and implemented in most statistical software.[13]

There are contexts where developing a plausible model is infeasible due to strict dependencies amongst covariates which impose constraints, e.g. total time spent active must be zero if the number of activities is zero. In these cases, the strategies implemented in common software packages will often impute implausible values, inappropriately assume MCAR and introduce bias,[14] or use single imputation approaches not accounting for uncertainty of imputed values and also potentially introducing bias.[9,14] An alternative is to use logic-based multiple imputation approaches to borrow information from observed data that already respects the imposed constraints.[15]

**Random hot deck imputation**

Random hot deck imputation is a logic-based approach that uses rules to build sets of candidates for matching rather than formal conditional models to predict missing values. First, one identifies a pool of "donors" with similar characteristics (covariates) to the record with missing data but with observed values. One then randomly selects a donor and imputes their observed value.[15] By using a random mechanism for imputation, random hot deck imputation can incorporate the additional uncertainty surrounding missing values.[15] A similar method that combines model- and logic-based approaches is fully conditional specification, where missing variables are imputed one at a time using regression models with constraints specific for that variable.[16] However, it is difficult to apply to longitudinal studies due to the larger number of constraints that must be incorporated into each regression model.[16] Incorporating constraints into random hot deck imputation is simpler and may be better suited for longitudinal data.[15]

While random hot deck imputation has commonly been used in surveys, several extensions to longitudinal data have been described in the literature. Little et al. and Wang et al. applied this method to multiply impute gaps in recurrent event data, using menstrual patterns as an example.[17,18] In this article, we provide a tutorial for applying random hot deck imputation[15] to missing longitudinal data. Unlike previous extensions, we restrict the donor pool to entries from the individual with missing data where appropriate.

Our approach is summarized by the following steps:

1. Identify covariates that are related to the missing variable and the nature of their relationship, including possible constraints.
2. Generate a donor pool by matching the record with missing data to other records based on observed covariate relationships.
3. Derive sampling probabilities for replacement values based on the observed data for records within the pool, sample, and impute a replacement value

This approach uses the same data to impute missing values as a model-based approach would use. In a model-based approach, one tries to encode domain knowledge about the relationships between variables in a joint probability model or set of conditional probability models so that data are MCAR conditional on the covariates in the model. Here, we use domain knowledge to create hierarchical strata, also assuming MCAR conditional on the matching covariates. By resampling values within the strata under these conditions, random hot deck imputation produces consistent estimates conditional on all relevant information.[15] Choosing the number of imputations also follows the same principles as model-based approaches.[19]

The randomness in the approach also allows one to compute confidence intervals using standard combining rules that account for variations within and between datasets. However, when there is a large percentage of missing information, standard methods underestimate variance because the same donor pool is used for all imputed datasets.[15] In this context, Bayesian Bootstrap or Adjusted Bayesian Bootstrap can be used to obtain appropriate confidence interval coverage.[15,20]

While we recommend researchers use model-based multiple imputation where possible, random hot deck imputation can provide less biased results and more appropriate levels of uncertainty than complete case analysis or single imputation methods when a model-based approach is infeasible.

## Illustrative Example

In this section, we illustrate our rationale for and implementation of random hot deck imputation using the CHAMPS-DK study.[3]

### Overview of data

We focus on weekly data collected via SMS on children's pain and sport participation. If no response was received for a question, the next question was not asked. As SMS messages were sent in a free-text field, responses did not always contain clear answers for the variable of interest. Where possible, entries were corrected by deduction, or else coded as missing.

**Pain**

Parents received an automated message asking whether their child experienced pain in their upper extremity, lower extremity, and spine in the past week, and whether pain was associated with a new injury (new pain) or continuing from a previous injury (old pain). These responses were converted into a composite variable (no pain, new pain in at least one body location, old pain in at least one body location and no new pain).

**Activity frequency**

After responding about pain, parents were asked to indicate the number of organized activity sessions (1-7, with 8 representing 8 or more) the child partook in outside of school that week.

**Types of sports**

After responding about frequency, parents were asked to indicate which sports were played in these sessions, with 1-9 representing different sports, and 10 representing "Other". We refer to the number of times a child played each sport in a week as the "sport count".

**Sport counts**

How parents indicated which sports were played sometimes resulted in missing data on sport counts. Consider a child with an activity frequency of 4 because they played football (code 1) three times and handball (code 2) once. Parents might answer 1112, providing one sport for each activity session, with no missing sport count data. However, other parents might answer 12 without specifying how many sessions of each sport were played, resulting in missing sport counts.

## Rationale for random hot deck imputation

While we imputed pain, frequency, sport, and sport counts using random hot deck principles, we focus on frequency, sport, and sport count data for brevity. There are multiple constraints amongst these variables that present challenges for model-based approaches. The activity frequency must be greater than or equal to the number of sports; each sport played must have an integer-valued sport count greater than 0; each sport not played must have a sport count of 0; and the sum of sport counts must total the frequency.

Because we were unable to develop a model that incorporated all constraints using standard imputation software, we applied random hot deck imputation to multiply impute missing data. Table 1 summarizes our approach.

**Table 1. Summary of random hot deck approach for frequency, sport, and sport count variables in the Childhood Health, Activity, and Motor Performance School Study Denmark (CHAMPS-DK) study.**

| Variable | Covariates | Constraints | Donor Pool | Sampling Method | Alternative Options |
|---|---|---|---|---|---|
| Frequency | • Pain<br>• Individual characteristics (e.g. gender, general level of activity)<br>• Age<br>• Seasonality<br>• Gender<br>• External factors (e.g. school events, weather) | • N/A | • Match on pain *within* individuals in nearby weeks (accounts for within individual characteristics, age, and seasonality) | • Sample difference between individual frequency and gender-specific median class frequency from donor pool<br>• Add difference to the median class frequency for the missing week (accounts for age, gender, external factors) | • Sample frequency directly from donor pool (does not account for external factors) |
| Sport | • Individual characteristics (e.g. gender, sport preference)<br>• Age<br>• Seasonality<br>• Frequency | • Number of sports cannot be greater than frequency | • Match on nearest frequency *within* individuals in nearby weeks (accounts for individual characteristics, age, and seasonality) | • Sample sport from donor pool<br>• If number of sampled sports is greater than the frequency, sample with replacement an equal number of sports as the frequency based on their relative proportion in nearby weeks (accounts for | • Additionally match on pain (reduces number of matching records within chosen time frame) |
| | | | | seasonality and frequency) | |
| Sport Count | • Individual characteristics (e.g. gender, sport preference)<br>• Age<br>• Seasonality<br>• Sport<br>• Frequency | • Number of sport counts must equal frequency<br>• Sport counts must be >0 for all sports played<br>• Sport counts must be 0 for all sports not played | • All nearby weeks *within* individuals where at least one of the sports were played (accounts for individual characteristics, age, seasonality, and sport) | • Calculate relative proportion of each sport in nearby weeks<br>• Sample with replacement an equal number of sports as the frequency based on their relative proportions (accounts for frequency) | • Assume some sports are more likely to be played in the same week (more complex logic) |

While we focus on sport participation, our approach can be applied to any context with constraints between variables. For example, activity frequency is akin to the total number of medications in a pharmacoepidemiology study. Sports are akin to drug types, and sport counts to side effects. The number of drug types cannot exceed the total number of medications. Probabilities of various side effects differ between drugs, and some side effects may never occur for certain drugs.

**Imputing frequency**

**1. Identify covariates**

One's activity frequency in a particular week is likely influenced by the presence or absence of pain. Additionally, activity frequency tends to change with season and age, and is likely more similar in nearby weeks than weeks further away. Further, children in the same class at school are exposed to similar factors (e.g. weather, school events) that may lead to individuals of the same class and gender being more active in certain weeks. To account for these external factors, we considered gender-specific median class frequencies in the missing and nearby weeks as covariates.

**2. Generate a donor pool**

We generated our donor pool by matching within individuals on pain in nearby weeks (we chose 7 weeks before and after the missing week) (Figure 1a-b). We used our composite pain variable with 3 levels (no pain, new pain, old pain), assuming pain in different body locations have the same effects on activity frequency.

When no matches were available in the 7 weeks before and after (3-month period), we extended the pool to include entries 12 weeks before and after (6-month period), 25 weeks before and after (1-year period), then the entire study. Others might choose smaller time windows when covariate relationships are sensitive to changes over time. Occasionally, no matches existed because the individual did not have any other entries with a particular pain value (new or old). However, they may have had entries with the other pain value. In these cases, we matched on any pain (new or old) in the 7 weeks before and after, and so forth. In cases where the individual had no entries with pain except the missing entry, we sampled from all entries in the 7 weeks before and after, even though these weeks had no pain.

### 3. Generate sampling probabilities and sample

Each week in the pool has an equal probability of being sampled. We randomly sampled one week from the pool (Figure 1c). We first imputed the difference between the individual's frequency and their gender-specific median class frequency from the sampled week as a measure of how much activity they did relative to their peers in the missing week. The imputed frequency for the missing week was the sum of the gender-specific median class frequency in the missing week and this difference (Figure 1d).

Our procedure is similar to random generation of values from a fixed effects model. We estimate the fixed effect for activity frequency from the observed data (the median frequency for individuals of the same class and gender, whom we assume come from the same distribution of relevant background characteristics). We sample a residual (the difference between the individual's frequency and the median frequency for their class and gender) from the pool of potential matches. The imputed value is then the sum of the fixed effect and the residual.

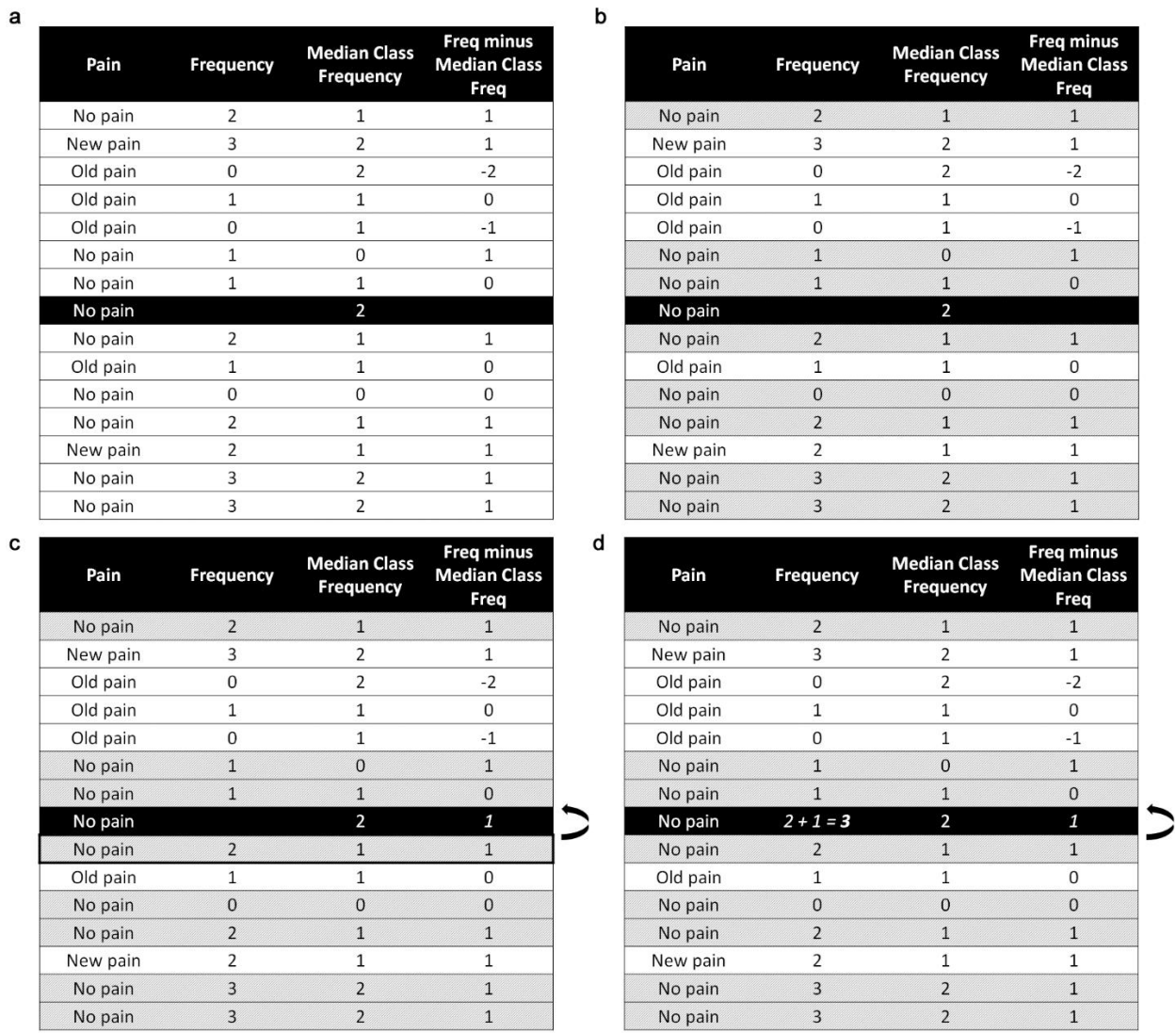

**Figure 1** Imputation of activity frequency. (**a**) There is one week where frequency is missing (black row). Pain is coded as no pain in any location (No pain), new pain in at least one location (New pain), and old pain in at least one location but no new pain (Old pain). The individual had no pain in this week. The median frequency for the individual's class and gender is calculated for the missing and surrounding weeks (Median Class Frequency). We also calculate the difference between the individual's frequency and the median frequency for all observed weeks (Freq minus Median Class Freq) as a measure of how much activity the individual does relative to their class and gender. (**b**) We match on nearby weeks with the same level of pain (gray rows). The sampling pool is comprised of eight weeks where the individual also experienced no pain. (**c**) One of the weeks in the sampling pool is randomly selected (outlined in black). The difference between the individual's frequency and the median class frequency for the sampled week is 1. This difference is imputed for the missing week. (**d**) The imputed frequency for the missing week is the sum of the median class frequency for the missing week and the imputed difference between the individual and median class frequency. In this example, the imputed difference of 1 is added to the median class frequency of 2 to obtain an imputed frequency of 3.

## Imputing sport

**1. Identify covariates**

We assume that while individuals are likely to play similar sports in nearby weeks, these sports might change over time and season. If an individual never played a particular sport, we assume that sport was not played in a missing week. We also assume that individuals played similar sports with similar frequencies in nearby weeks.

**2. Generate donor pool**

We generated our donor pool by matching within individuals on closest frequency to the missing week within nearby weeks (7 weeks before and after) (Figure 2a-b). We did not extend the time window to match on exact frequency because we assume sports participation differs by season. This is an example where constraints in our data make model-based approaches difficult.

If there were no matches on closest frequency (i.e. if an individual did not have any observed frequency in nearby weeks or all frequencies were 0), we extended the time window to 12 weeks before and after (6-month period), 25 weeks (1-year period), then the entire study.

**3. Generate sampling probabilities and sample**

Each week in the pool had an equal probability of being sampled. If all records in the donor pool had the same frequency as the missing week, we randomly sampled one of these weeks (Figure 2c) and imputed its sports (Figure 2d).

Sometimes the donor pool contained records with frequencies lower or greater than the missing week. If the sampled week had a lower frequency than the missing week, we imputed its sports. These records would still be missing sport counts (number of times each sport was played). We describe how we imputed sport counts in Section 2.5.

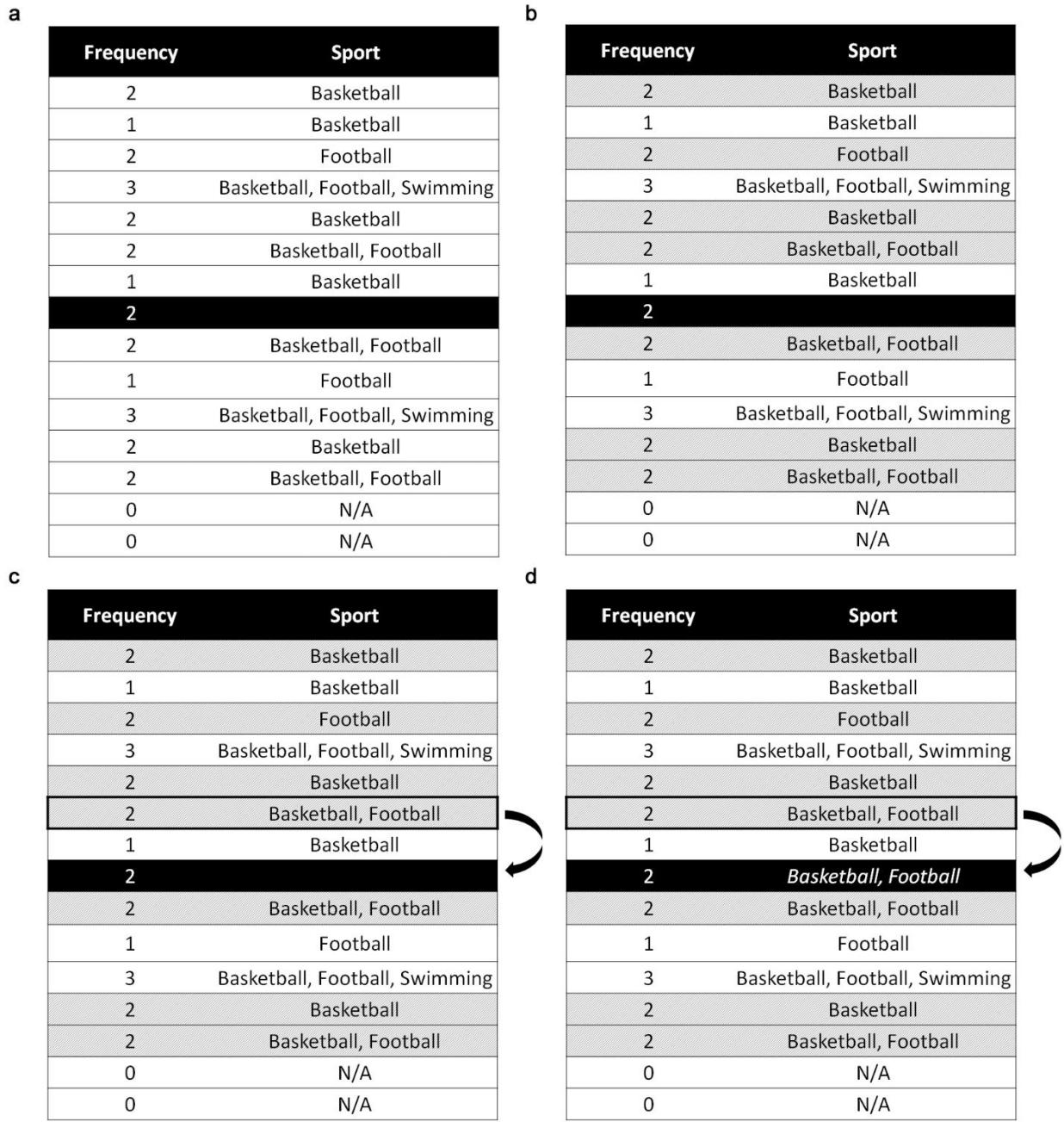

**Figure 2** Imputation of sport. (**a**) There is one week where the sports performed are missing (black row). The individual had a total activity frequency of 2 in this week. (**b**) We match on closest frequency in the nearby weeks. The sampling pool is comprised of weeks where the individual also had frequencies of 2 (gray rows). (**c**) One of these weeks is randomly sampled

with equal probability (outlined in black). (**d**) The sports from the sampled week are imputed for the missing week.

If the sampled week had a frequency greater than the missing week, two possibilities existed. If the number of sports was equal or less than the frequency of the missing week, we imputed the sampled week's sports. However, if the number of sports was greater than the frequency of the missing week, imputing all the sports would break our constraints. In Figure 3a, the missing week has a frequency of 2, but nearby weeks have frequencies of 3 or more. The sampled week (Figure 3b) has three sports. To determine which sports to impute, we calculated sampling probabilities for each of the three sports according to their relative proportion in nearby weeks and sampled two sports with replacement (Figure 3c-d).

Our approach assumes that the sports one plays were only related to the individual's activity frequency that week and sports played in nearby weeks (i.e. seasonality). More complex matching constraints could include the presence of pain. More constraints will reduce the number of matching records within the chosen time frame.

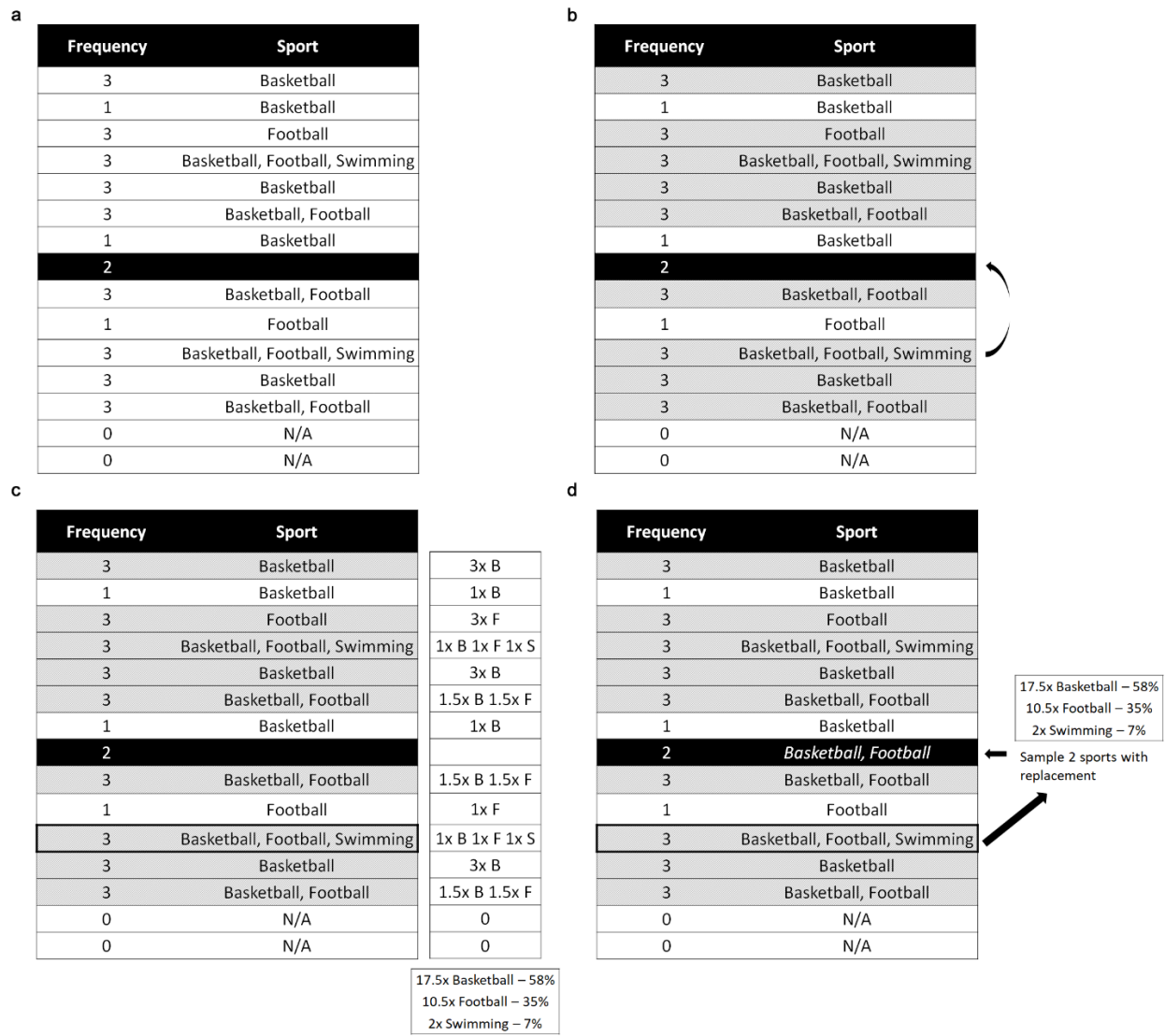

**Figure 3** Imputation of sport where the number of sports is greater than the frequency. (**a**) There is one week where the sports performed are missing (black row). The individual had a total activity frequency of 2 in this week. (**b**) The sampling pool is comprised of nearby weeks with the closest frequency to the missing week. Since there are no weeks with a frequency of 2, we match on weeks with frequencies of 3 (gray rows). One of these weeks is randomly sampled. (**c**) The sampled week has 3 sports, while the missing week only has a frequency of 2. The number of times in nearby weeks that the individual participated in each sport is determined. For weeks where the frequency is greater than the number of sports, the frequency is divided equally. The relative amount that the individual participated in each sport in nearby weeks is used as the sampling probability. Since the individual did basketball 10.5 times, football 7.5 times, and swimming 1 time, the probabilities are 55% (10.5/19), 40% (7.5/19), and 5% (1/19) respectively. (**d**) Sports are randomly sampled using the sampling probabilities and imputed for the missing week. Basketball and football are randomly imputed.

**Imputing sport counts**

In this study, most parents indicated which sports were played but not how many times each sport was played (e.g. Figure 4a, frequency "3", sport "Basketball, Football"). We imputed sport counts in cases where the total frequency (observed or imputed) was greater than the number of sports. For each week with missing sport counts, each listed sport was played at least once. Therefore, the number of missing counts is the total frequency minus the number of sports (Figure 4b).

**1. Identify covariates**

Similar to activity frequency and sports participation, we believe the relative frequency with which individuals participate in different sports differs by age, gender and season. Individuals are likely to have similar relative frequencies in nearby weeks. Because individuals participate in different sports at different frequencies, we only borrowed information from within individuals.

**2. Generate donor pool**

Our donor pool included all nearby weeks (7 weeks before and after) for the individual with missing sport counts that included sports from the missing week. Because sports that were not played must have a zero probability for the remaining counts, other weeks were not included in the pool.

**3. Generate sampling probabilities and sample**

Probabilities were derived by dividing the total counts for each sport in nearby weeks by the total frequency that the sports in the missing week were played in nearby weeks (Figure 4c). In our example, we know the child played basketball and football once each. In nearby weeks, they played basketball 18 times and football 8 times (69% basketball; 31% football). We used these probabilities to sample the remaining sport count so that the total sport count matched the frequency (Figure 4d).

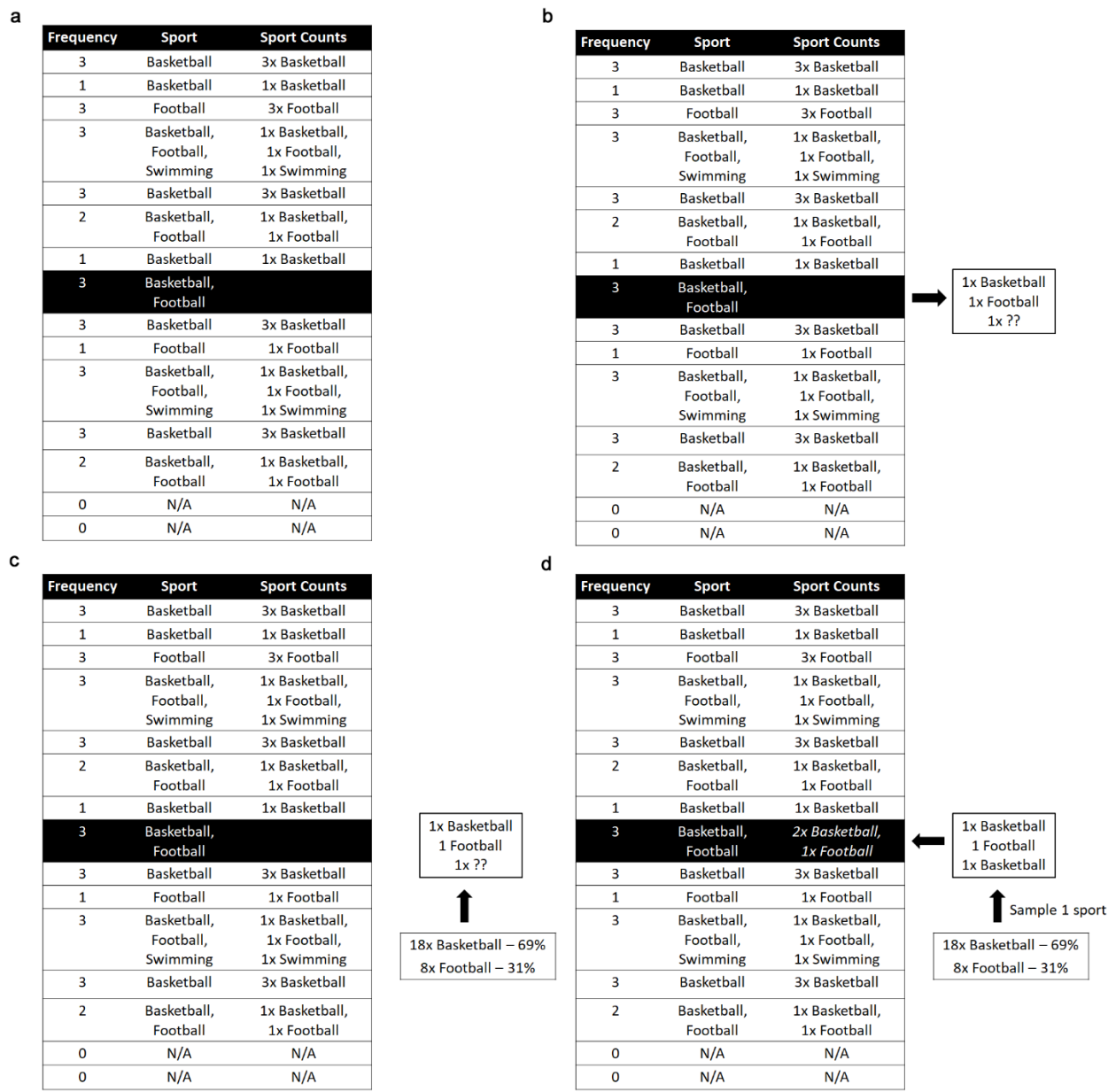

**Figure 4** Imputation of sport counts where a single week is missing. (**a**) There is one week where the total frequency is greater than the number of sports performed (black row). We would like to impute individual counts for each sport that was done. (**b**) The individual participated in at least one session of basketball and one session of football. As the total frequency for the missing week is 3, we still need to impute a single count that is either basketball or football. (**c**) The relative proportion of each sport in the sampling pool (i.e. the sports that were done in the missing week;

basketball and football) is calculated for the nearby weeks and used as the sampling probabilities. As basketball was done 9 times and football was done 5 times, the probabilities are 64% (9/14) and 36% (5/14) respectively. (**d**) Basketball is randomly sampled. Sport counts are imputed as two sessions of basketball and one session of football.

Sometimes, sport counts were also missing for nearby weeks (Figure 5a). For these weeks, we divided their frequency by the number of sports to obtain average counts (Figure 5b). These counts were not imputed into the main dataset; rather, they were temporarily used to determine the probabilities for the week of interest. The counts from observed weeks and average counts for missing weeks were then summed as above and divided by the total frequency to obtain sampling probabilities (Figure 5c-d). Once data were imputed for the first missing week, we proceeded to the next missing week.

Our approach assumes that the probability of playing a particular sport is independent of the other sports. Alternatively, we could use more complex logic that assumes some sports are more likely to be played together, and adjust our sampling scheme accordingly.

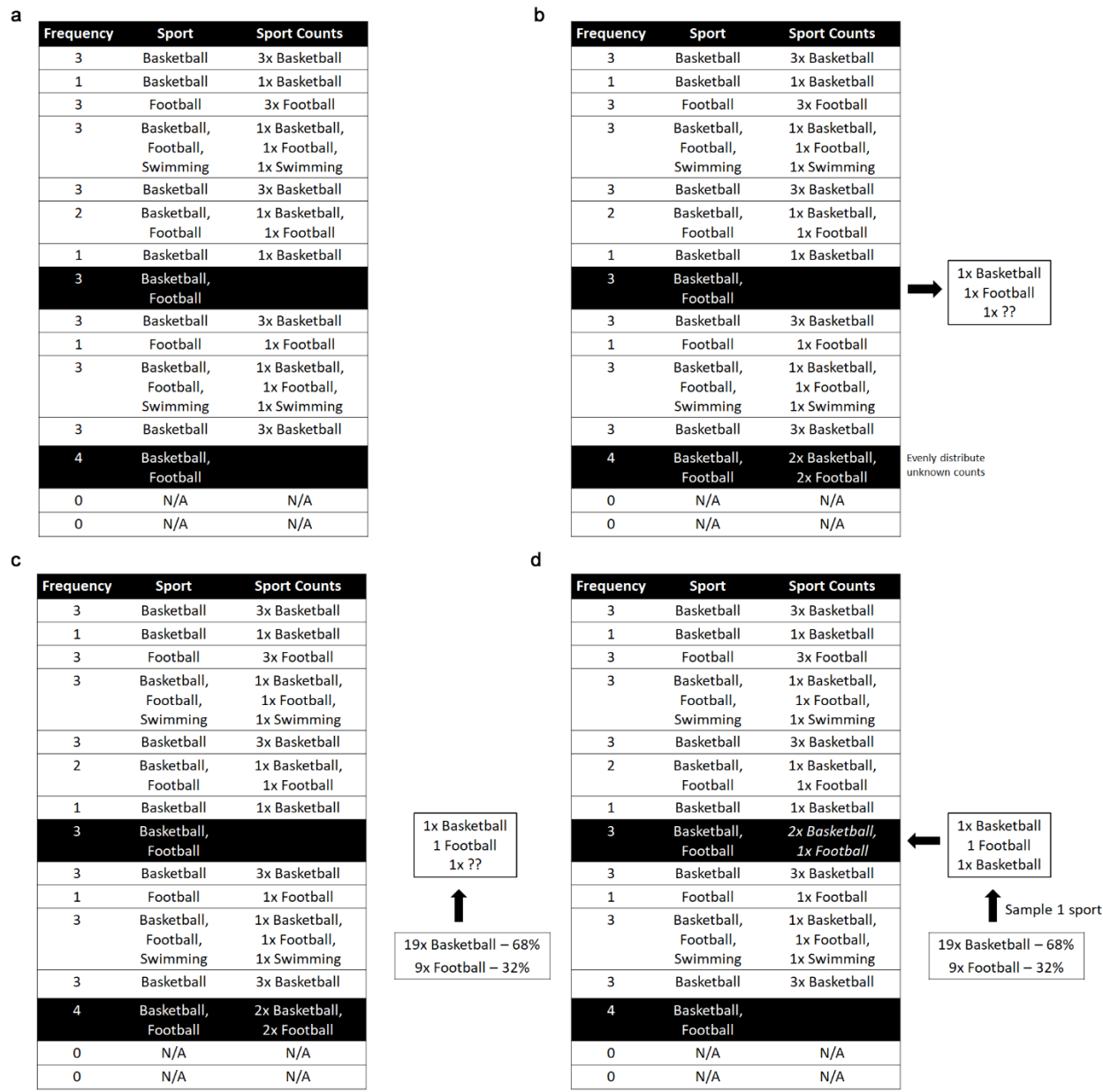

**Figure 5** Imputation of sport counts where multiple weeks are missing. (**a**) There are two weeks where the total frequency is greater than the number of sports (black rows). We would like to impute individual counts for each sport that was done for both weeks. We focus on imputing sport counts for the first missing week. (**b**) In the missing week, the individual had a frequency of 3 and participated in basketball and football. They must have participated in one session each of basketball and football. We must therefore impute a single count. Sport counts are calculated

for the nearby weeks. When the frequency is greater than the number of sports (i.e. for the week with frequency 4), the counts are evenly distributed (assigning 2 counts to basketball and 2 counts to football). (**c**) The relative proportion of each sport in the sampling pool (i.e. matching the sports that were done in the missing week) is calculated for the nearby weeks and used as the sampling probabilities. Since basketball was done 10 times and football 7 times, the sampling probabilities are 59% (10/17) and 41% (7/17) respectively. (**d**) Basketball is randomly sampled. Sport counts are imputed as 2 sessions of basketball and 1 session of football.

# Conclusion

Model-based multiple imputation may result in implausible values where there are constraints between variables. Random hot deck multiple imputation is a possible alternative for longitudinal data that respects constraints between variables by creating a pool of matching records from observed data, generating probabilities for these records, and randomly sampling an imputed value. Although this approach requires many assumptions, any model-based approach would have to include the same assumptions or risk imputing implausible values.